\documentstyle[prl,eqsecnum,aps]{revtex}
\begin{document}
\author{Jian-Qi Shen \footnote{E-mail address: jqshen@coer.zju.edu.cn}}
\address{1. Zhejiang Institute of Modern Physics and Department of
Physics\\ 2. Centre for Optical and Electromagnetic Research\\
Zhejiang University, Hangzhou 310027, P. R. China}
\date{\today }
\title{Dynamics of Gravitomagnetic Charge \footnote{The original version of this paper was written in 1999$\sim $2000. In the present paper I would like to discuss some aspects of gravitomagnetic charge in more detail. It won't be submitted elsewhere for publication since most of the subjects in this manuscript are analogous to those presented in one of my papers published in 2002 $\left[ Gen. Rel. Gra. {\bf 34}(9),1423-1435 (2002)\right] $. Readers may also be referred to this published paper for a brief review of the essential features of the potential field equations of the hypothetical gravitomagnetic monopole.}}
\maketitle

\begin{abstract}
The physically interesting gravitational analogue of magnetic monopole in
electrodynamics is considered in the present paper. The author investigates
the field equation of gravitomagnetic matter, and the exact static
cylindrically symmetric solution of field equation as well as the motion of
gravitomagnetic charge in gravitational fields. Use is made of the mechanism
of gravitational Meissner effect, a potential interpretation of anomalous,
constant, acceleration acting on the Pioneer 10/11, Galileo and Ulysses
spacecrafts is also suggested.

PACS Ref: 04.20.-q, 04.20.Fy, 04.25.-g
\end{abstract}

\pacs{PACC:}

\section{Introduction}

It is well known that the field equation of general relativity
under low-motion weak-field approximation is analogous to
Maxwell$^{,}$s equation of electromagnetic field. This similarity
leads us to consider the following interesting gravitational
analogues of electromagnetic phenomena: (i) in electrodynamics a
charged particle is acted upon by Lorentz magnetic force, in the
same manner, a particle is also acted upon by the gravitational
Lorentz force in weak-gravity theory \cite{Kleinert,Shen}.
According to the principle of equivalence, further analysis shows
that in the non-inertial rotating reference frame, this
gravitational Lorentz force is just the fictitious Coriolis
force\cite{Shen2}; (ii) there exists the Aharonov-Bohm effect in
electrodynamics\cite{Bohm}, accordingly, the so-called
gravitational Aharonov-Bohm effect, {\it i.e.}, the gravitational
analogue of Aharonov-Bohm effect also exists in theory of
gravitation, which is now termed Aharonov-Carmi
effect\cite{Aharonov,Anandan,Dresden}; (iii) a particle with
intrinsic spin possesses a gravitomagnetic moment of such
magnitude that it equals the spin of this particle which can be
coupled to gravitomagnetic fields. Mashhoon$^{,}$s spin-rotation
coupling is the non-inertial realization of the interaction of
gravitomagnetic moment with gravitomagnetic
field\cite{Mashhoon1,Mashhoon2}. We think that investigation of
general characters of gauge field deserves generalization to that
of the gravitational field\cite{Hogan,Ingraham}. In this paper, we
study another physically interesting phenomenon, {\it i.e.}, the
gravitational analogue of magnetic charge in electrodynamics.

In electrodynamics, electric charge is a Noether charge while its
dual charge ( magnetic charge ) is a topological charge, since the
latter is related to the singularity of non-analytical magnetic
vector potentials. Magnetic monopole\cite{Dirac} attracts
attentions of many physicists in various fields such as gauge
field theory, grand unified theory, particle physics and
cosmology\cite{Schwinger,Yang1,Yang2,Hooft,Tchra}. In the similar
fashion, it is also interesting to consider gravitomagnetic charge
which is the source of gravitomagnetic field just as mass (
gravitoelectric charge ) is the source of gravitoelectric field (
Newtonian gravitational field ). In this sense, gravitomagnetic
charge is also termed dual mass. It should be noted that the
concept of mass is of no significance for the gravitomagnetic
charge, and then it is of interest to investigate the relativistic
dynamics and gravitational effects of this topological dual mass.
Historically, gravitomagnetic monopole attracts attentions of many
authors in both classical dynamics and quantization
formulation\cite{Taub,Newman,Dowker,Nouri,Dowker2,Zee}. For
historical review one may be referred to the paper by Linden-Bell
and Nouri-Zonoz\cite{Nouri2}.

In literature, to the best of our knowledge, many authors were
concerned with {\it monopole} rather than with {\it charge},
namely, they considered equations of motion of gravitomagnetic
monopoles more than field equations of gravitomagnetic
charges/matter. In this paper, the author proposes the field
equation of gravitomagnetic matter and investigates the equation
of motion of gravitomagnetic monopole, and then exactly solves the
static cylindrically symmetric solution and discusses the motion
of a photon in the gravitomagnetic fields. In Sec. V, a potential
interpretation of anomalous, constant, acceleration acting on the
Pioneer 10/11, Galileo and Ulysses spacecraft\cite{Anderson} is
suggested. The relation between the non-analytic property of
metric and the topological gravitomagnetic charge is discussed in
Sec. VI. In Sec. VII, the author concludes with some remarks.

\section{Topological charge and field equation of gravitomagnetic matter}

First we take into consideration the $SO(3)$ non-Abel magnetic
monopole proposed by Li {\it et al.}\cite{Li1,Li2}, which enables
one to consider the topological dual mass in curved spacetime,
then we suggest the gravitational field equation of
gravitomagnetic charge.

The linear element on the two-dimensional spherical surface is ${\rm d}%
s^{2}=r^{2}{\rm d}\theta ^{2}+r^{2}\sin ^{2}\theta {\rm d}\varphi ^{2}$
where the non-vanishing Christoffel symbols $\Gamma _{\mu j}^{i}$ $(\mu
,i,j=1,2)$ is given as follows
\begin{equation}
\Gamma _{21}^{2}=\Gamma _{12}^{2}=\frac{\cos \theta }{\sin \theta },\quad
\Gamma _{22}^{1}=-\sin \theta \cos \theta  \label{eq1}
\end{equation}%
and its matrix form $(\Gamma _{\theta })_{j}^{i}$ and $(\Gamma _{\varphi
})_{j}^{i}$ are therefore
\begin{equation}
(\Gamma _{\theta })=\left(
\begin{array}{cc}
0 & 0 \\
0 & \frac{\cos \theta }{\sin \theta }%
\end{array}%
\right) ,\quad (\Gamma _{\varphi })=\left(
\begin{array}{cc}
0 & -\sin \theta \cos \theta \\
\frac{\cos \theta }{\sin \theta } & 0%
\end{array}%
\right) .  \label{eq2}
\end{equation}%
Under the transformation $M^{-1},$ the basic vector $r\sin \theta {\rm d}%
\varphi $ that varies with $\theta $ may be transformed into what is $\theta
$-independent, namely,
\begin{equation}
\left(
\begin{array}{c}
{\rm d}\theta \\
{\rm d}\varphi%
\end{array}%
\right) =M^{-1}\left(
\begin{array}{c}
r{\rm d}\theta \\
r\sin \theta {\rm d}\varphi%
\end{array}%
\right) ,\quad M=\left(
\begin{array}{cc}
r & 0 \\
0 & r\sin \theta%
\end{array}%
\right) .  \label{eq3}
\end{equation}%
Accordingly, $\Gamma _{\mu }$ is transformed into $\tilde{\Gamma}_{\mu
}=M\Gamma _{\mu }M^{-1}+M\partial _{\mu }M^{-1}$ and the results are of the
form
\begin{equation}
\tilde{\Gamma}_{\theta }=0,\quad \tilde{\Gamma}_{\varphi }=\left(
\begin{array}{cc}
0 & -\cos \theta \\
\cos \theta & 0%
\end{array}%
\right) =-\cos \theta X_{3}  \label{eq4}
\end{equation}%
with $X_{3}=\left(
\begin{array}{cc}
0 & 1 \\
-1 & 0%
\end{array}%
\right) $ being one of the generators of $SO(3)$ group. There exists the
connection between $\tilde{\Gamma}_{\mu }$ and the potential $W_{\mu }^{3}$
which may be written as $\tilde{\Gamma}_{\mu }=gW_{\mu }^{3}X_{3}$ and we
therefore obtain $gW_{\theta }^{3}=0,\quad gW_{\varphi }^{3}=-\cos \theta $,
and the non-vanishing field strengths are written in the forms
\begin{equation}
f_{\theta \varphi }^{3}=-f_{\varphi \theta }^{3}=\frac{\partial W_{\varphi
}^{3}}{\partial \theta }-\frac{\partial W_{\theta }^{3}}{\partial \varphi }=%
\frac{\sin \theta }{g}.  \label{eq5}
\end{equation}%
It follows that the source charge $\tilde{g}$ of field strength $\ f_{\mu
\nu }^{3}$ is expressed by
\begin{equation}
\tilde{g}=\oint f_{\theta \varphi }^{3}{\rm d}\theta {\rm d}\varphi =\frac{%
4\pi }{g}.  \label{eq6}
\end{equation}%
It should be noted that $g$ is Noether charge while $\tilde{g}$ is
topological charge whose existence is related to the singularities
of field potentials. The work of Li {\it et al.}\cite{Li1} leads
us to consider the dual charge associated with spacetime. In fact,
although no topological charge $\tilde{g}$ exists seen from in the
three-dimensional space, the reason why the two-dimensional
surface is curved seen from the observer on the above
two-dimensional spherical surface can be equivalently ascribed to
the presence of this kind of dual charge. The concept of above
topological charge may be readily generalized to that of
gravitomagnetic charge.

The dual of Riemann curvature tensor, $R_{\mu \nu \gamma \delta },$ may be
defined as follows
\begin{equation}
\tilde{R}_{\mu \nu \gamma \delta }=\epsilon _{\mu \nu }^{\quad \alpha \beta
}R_{\gamma \delta \alpha \beta }+\epsilon _{\gamma \delta }^{\quad \alpha
\beta }R_{\mu \nu \alpha \beta }  \label{eq7}
\end{equation}%
with $\epsilon _{\mu \nu }^{\quad \alpha \beta }$ being the completely
antisymmetric Levi-Civita tensor satisfying its covariant derivative $%
\epsilon _{\mu \nu \quad ;\xi }^{\quad \alpha \beta }=0$. It
should be noted that $\tilde{R}_{\mu \nu \gamma \delta }\equiv 0$
in the absence of gravitomagnetic matter since no singularities
associated with topological charge exist in the metric functions
and therefore Ricci identity still holds. However, once the metric
functions possess non-analytic properties in the presence of
gravitomagnetic matter ( should such exist ), the dual curvature
tensor is therefore no longer vanishing due to the violation of
Ricci identity. The dual curvature scalar that contains only the
first derivative and second derivative of field $g_{\mu \nu }$ can
be chosen to be $\tilde{R}=-\epsilon ^{\mu \nu \alpha \beta
}R_{\mu \nu \alpha \beta }$. By making use of the variational
principle, the dual Einstein$^{,}$s tensor is obtained as follows
\begin{eqnarray}
\delta \int_{\Omega }-\sqrt{-g}\epsilon ^{\mu \nu \alpha \beta
}R_{\mu \nu
\alpha \beta }{\rm d}\Omega &=&\frac{1}{2}\int_{\Omega }\sqrt{-g}(\tilde{R}%
_{\mu \nu }-2\epsilon _{\quad \mu }^{\delta \quad \ \alpha \beta
}R_{\nu \delta \alpha \beta })\delta g^{\mu \nu }{\rm d}\Omega
\nonumber \\ &=&\int_{\Omega }\sqrt{-g}\tilde{G}_{\mu \nu }\delta
g^{\mu \nu }{\rm d}\Omega
   \label{eq8}
\end{eqnarray}
with $\tilde{R}_{\mu \nu }=g^{\sigma \tau }\tilde{R}_{\sigma \mu
\nu \tau }.$ Further calculation yields
\begin{equation}
\tilde{G}_{\mu \nu }=\frac{1}{2}(\epsilon _{\nu }^{\quad \alpha \beta \gamma
}R_{\mu \alpha \beta \gamma }-\epsilon _{\mu }^{\quad \alpha \beta \gamma
}R_{\nu \alpha \beta \gamma }),  \label{eq9}
\end{equation}%
which is considered the dual of the Einstein$^{,}$s tensor arising on the
left handed side of gravitational field equation of gravitomagnetic matter.
Note, however, that once the gravitomagnetic charge is absent in spacetime, $%
\tilde{G}_{\mu \nu }$ vanishes due to the Ricci identity. But, the
non-analytic metric functions caused by the presence of gravitomagnetic
charge may result in the violation of Ricci identity and therefore lead to $%
\tilde{G}_{\mu \nu }\neq 0$. Since the dual Einstein's tensor is
an antisymmetric tensor, in what follows we will construct the
antisymmetric source tensor of gravitomagnetic charge that is on
the right handed side of gravitational field equation. We show
that for the Fermion field, the antisymmetric tensors constructed
in terms of field $\psi $ and space-time derivatives are of the
form

\begin{equation}
K_{\mu \nu }=i\bar{\psi}(\gamma _{\mu }\partial _{\nu }-\gamma _{\upsilon
}\partial _{\mu })\psi ,\quad H_{\mu \nu }=\epsilon _{\mu \nu }^{\quad
\alpha \beta }K_{\alpha \beta }  \label{eq10}
\end{equation}%
and regard the linear combination of them as the source tensor in
the field equation of gravitomagnetic charge, where $\gamma _{\mu
} '$s
denote the general Dirac matrices with respect to $x^{\mu }$ and satisfy $%
\gamma _{\mu }\gamma _{v}+\gamma _{\nu }\gamma _{\mu }=2g_{\mu \nu }.$ Thus
the field equation governing the distribution of gravitation of
gravitomagnetic charge may be given as follows
\begin{equation}
\tilde{G}_{\mu \nu }=\kappa _{1}K_{\mu \nu }+\kappa _{2}H_{\mu \nu }
\label{eq11}
\end{equation}%
with $\kappa _{1},\kappa _{2}$ being the coupling coefficients
between gravitomagnetic matter and gravity. One of the advantages
of this field equation is that it does not introduce extra tensor
potentials when allowing for gravitomagnetic monopole densities
and currents. This fact is in analogy with that in
electrodynamics, where the equation $\partial _{\nu
}\tilde{F}^{\mu \nu }=J_{\rm \bf M}^{\mu}$ governs the motion of
electromagnetic fields produced by magnetic monopoles.
Additionally, further analysis shows
that the linear combination of the following antisymmetric tensors, $i\bar{%
\psi}(\gamma _{\mu }\gamma _{\nu }-\gamma _{\upsilon }\gamma _{\mu })\psi $
and $i\epsilon _{\mu \nu }^{\quad \alpha \beta }\bar{\psi}(\gamma _{\alpha
}\gamma _{\beta }-\gamma _{\beta }\gamma _{\alpha })\psi $ serves as the
{\sl cosmological term} in the above gravitational field equation of Fermion
field.

It is believed that there would exist formation ( and creation )
mechanism of gravitomagnetic charge in the gravitational
interaction, just as some prevalent theories\cite{Hooft} provide
the theoretical mechanism of existence of magnetic monopole in
various gauge interactions. Magnetic monopole in electrodynamics
and gauge field theory has been discussed and sought after for
decades, and the existence of the $^{,}$t Hooft-Polyakov monopole
solution has spurred new interest of both theorists and
experimentalists\cite{Hooft,Polyakov,Polyakov2}. Similar to
magnetic monopole, gravitomagnetic charge is believed to give rise
to such situations. If it is indeed present in universe, it will
also lead to significant consequences in astrophysics and
cosmology. We emphasize that although it is the classical solution
to the field equation as discussed above, this kind of topological
gravitomagnetic monopole may arise not as fundamental entities in
gravity theory.

\section{Motion of gravitomagnetic charge}

According to its gravitational properties, gravitomagnetic charge can be
called {\sl dual mass}, and accordingly, a point-like particle possessing
mass is also called gravitoelectric monopole. This, therefore, implies that
the concept of mass is of no essential significance for gravitomagnetic
matter; it is of interest to investigate the motion of gravitomagnetic
monopole in curved spacetime. Although Ricci identity is violated due to the
non-analytic properties caused by the existence of gravitomagnetic charge,
Bianchi identity still holds in the presence of gravitomagnetic charge. It
follows that the covariant divergence of $\tilde{G}^{\mu \nu }$ vanishes,
namely,
\begin{equation}
\tilde{G}_{\quad ;\nu }^{\mu \nu }=0.  \label{eq12}
\end{equation}%
Then in terms of the following field equation
\begin{equation}
\tilde{G}^{\mu \nu }=S^{\mu \nu }  \label{eq13}
\end{equation}%
with the antisymmetric source tensor of gravitomagnetic matter $S^{\mu \nu }$
being $\kappa _{1}K^{\mu \nu }+\kappa _{2}H^{\mu \nu }$, one can arrive at
\begin{equation}
S_{\quad ;\nu }^{\mu \nu }=0  \label{eq14}
\end{equation}%
which can be regarded as the equation of motion of gravitomagnetic charge in
the curved spacetime. It is useful to obtain the low-motion and
weak-field-approximation form of Eq. (\ref{eq14}), which enables us to
guarantee that Eq. (\ref{eq14}) is indeed the equation of motion of
gravitomagnetic monopole.

The general Dirac matrices under the weak-field approximation may be
obtained via the relations $\gamma _{\mu }\gamma _{v}+\gamma _{\nu }\gamma
_{\mu }=2g_{\mu \nu }$ and the results are given as follows
\begin{equation}
\gamma ^{0}=(1+g^{0})\beta ,\quad \gamma ^{i}=g^{i}\beta +\gamma
_{\rm M}^{i}, \label{eq15}
\end{equation}%
where $i=1,2,3;$ $\beta =\gamma _{\rm M}^{0}.$ $\gamma _{\rm
M}^{0}$ and $\gamma _{\rm M}^{i}$ are the constant Dirac matrices
in the flat Minkowski spacetime. The gravitoelectric potential is
defined to be $g^{0}=\frac{g^{00}-1}{2},$ and gravitomagnetic
vector potentials are $g^{i}=g^{0i}\quad (i=1,2,3).$ In the
framework of dynamics of point-like particle, the source tensor is
therefore rewritten as
\begin{equation}
S^{\mu \nu }=\rho \left[ \kappa _{1}(g^{\mu }U^{\nu }-g^{\nu }U^{\mu
})+\kappa _{2}\epsilon ^{\mu \nu \alpha \beta }(g_{\alpha }U_{\beta
}-g_{\beta }U_{\alpha })\right] ,  \label{eq16}
\end{equation}%
where $\rho $ denotes the density of gravitomagnetic matter. It
follows from Eq. (\ref{eq14}) and Eq. (\ref{eq16}) that there
exists the gravitational Lorentz force density in the expression
for the force acting on the gravitomagnetic charge, namely, by
ignoring some small term and using the static condition
$\frac{\partial }{\partial x^{0}}g^{0}=0$, one can obtain the
following expression ( in the unit $c=1$ )
\begin{eqnarray}
\kappa _{1}g^{0}\frac{\partial }{\partial x^{0}}{\bf v} &=&2\kappa
_{2}\left[{\nabla \times {\bf g}-{\bf v}\times\left ({\nabla g^{0}-\frac{\partial }{%
\partial x^{0}}{\bf g}}\right)}\right]-2\kappa _{2}{\bf g}\times \left({\frac{\partial }{%
\partial x^{0}}{\bf v}+\nabla g^{0}}\right)  \nonumber \\
&&+\kappa _{1}g^{0}\frac{\partial }{\partial x^{0}}{\bf g}-2\kappa
_{2}g^{0}\left({\nabla g^{0}\times {\bf v}}\right)  \label{eq170}
\end{eqnarray}%
with ${\bf v}$ being the velocity of the tested gravitomagnetic monopole. It
is apparent that $\nabla \times {\bf g}-{\bf v}\times (\nabla g^{0}-\frac{%
\partial }{\partial x^{0}}{\bf g})$ is the expression associated with
gravitational Lorentz force density. Note that in Eq. (\ref{eq170}) $\kappa
_{1},\kappa _{2}$ are considered coupling constants. However, further
analysis shows that at least one of them is not a constant and if the
relation
\begin{equation}
\kappa _{1}g^{0}=2\kappa _{2}  \label{eq18}
\end{equation}%
between them is assumed, then Eq. (\ref{eq170}) may be rewritten as
\begin{equation}
\frac{\partial }{\partial x^{0}}{\bf v}=\left[{\nabla \times {\bf
g}-{\bf v}\times \left({\nabla g^{0}-\frac{\partial }{\partial
x^{0}}{\bf g}}\right)}\right],  \label{eq19}
\end{equation}%
where we ignore the small term of second order and the derivative term of
coupling coefficients with respect to spacetime coordinate, $x^{\mu }$. It
is well known that the form of Eq. (\ref{eq19}) is the equation of motion of
a particle acted upon by the Lorentz force. Hence, Eq. (\ref{eq14}) is
believed to be the generally relativistic equation of motion of
gravitomagnetic monopole in the Riemann spacetime.

Investigation of relativistic dynamics of the topological dual mass is of
interest. It should be noted that although gravitomagnetic monopole does not
possess mass, it still has energy. Since the dual mass is a kind of
topological charge which is very different from Noether charge, no mechanism
of interaction may turn it into mass (gravitoelectric charge), and vice
versa.

The gravitational analogue of Meissner effect in superconductivity is
gravitational Meissner effect. Due to the conservation law of momentum,
mass-current density may be conserved in the process of scattering in
perfect fluid, which is analogous to the superconductivity of
superconducting electrons in superconductors cooled below $T_{c}.$ Since
gravitational field equation under linear approximation is similar to the
London$^{^{\prime }}$s equations of superconductivity, one can predict that
gravitational Meissner effect arises in perfect fluid. The author holds that
the investigation of both the effect of gravitomagnetic matter and
gravitational Meissner effect may provide us with a valuable insight into
the problem of cosmological constant and vacuum gravity\cite%
{Weinberg,Datta,Alvarenga,Cap}: the gravitoelectric field (
Newtonian field of gravity ) produced by the gravitoelectric
charge ( mass ) of the vacuum quantum fluctuations is exactly
cancelled by the gravitoelectric field due to the induced current
of the gravitomagnetic charge of the vacuum quantum fluctuations;
the gravitomagnetic field produced by the gravitomagnetic charge (
dual mass ) of the vacuum quantum fluctuations is exactly
cancelled by the gravitomagnetic field due to the induced current
of the gravitoelectric charge ( mass current ) of the vacuum
quantum fluctuations. Thus, at least in the framework of
weak-field approximation, the extreme space-time curvature of
vacuum caused by its large energy density does not arise, and the
gravitational effects of cosmological constant is eliminated by
the contributions of the gravitomagnetic charge ( dual mass ). If
gravitational Meissner effect is of really physical significance,
then it is necessary to apply this effect to the early universe.
Some related topics such as gravitational Hall effect and
gravitational magnetohydrodynamics may be considered for further
consideration.

\section{Static cylindrically symmetric exact solution of field equation}

In this section the static cylindrically symmetric gravitomagnetic field and
the evolution of wavefunction of photon in gravitomagnetic field are
considered. Suppose that the form of linear element describing the static
cylindrically symmetric gravitomagnetic field is given by
\begin{equation}
{\rm d}s^{2}={\rm d}( {x^{0}})^{2}-{\rm d}x^{2}-{\rm d}y^{2}-{\rm d}z^{2}+2g_{0x}(y)%
{\rm d}x^{0}{\rm d}x+2g_{0y}(x){\rm d}x^{0}{\rm d}y,  \label{eq20}
\end{equation}%
where we assume that the gravitomagnetic potentials $g_{0x}$ and $g_{0y}$
are functions with respect to $y$ and $x$, respectively. Thus we obtain all
the only nonvanishing values of Christoffel symbols as follows:
\begin{equation}
\Gamma _{0,xy}=\Gamma _{0,yx}=\frac{1}{2}\left({\frac{\partial g_{0x}}{\partial y}+%
\frac{\partial g_{0y}}{\partial x}}\right),\quad \Gamma
_{x,0y}=\Gamma
_{x,y0}=-\Gamma _{y,0x}=-\Gamma _{y,x0}=\frac{1}{2}\left({\frac{\partial g_{0x}}{%
\partial y}-\frac{\partial g_{0y}}{\partial x}}\right).  \label{eq21}
\end{equation}%
\ Since the field equation of gravitomagnetic matter is the antisymmetric
equation, we might as well take into account a simple case of the following
simplified equation
\begin{equation}
\epsilon ^{0\alpha \beta \gamma }R_{\quad \alpha \beta \gamma }^{0}=\rho _{M}
\label{eq22}
\end{equation}%
with $\rho _{M}$ being the parameter associated with the coupling constants
and gravitomagnetic charge. It is therefore apparent that Eq. (\ref{eq22})
agrees with Eq. (\ref{eq13}). Hence, the solution of the former equation
also satisfies the latter. For the reason of the completely antisymmetric
property of the Levi-Civita tensor, the contravariant indices $\alpha ,\beta
,\gamma $ should be respectively taken to be $x,y,z$, namely, we have
\begin{equation}
\epsilon ^{0\alpha \beta \gamma }R_{\quad \alpha \beta \gamma
}^{0}=2\epsilon ^{0xyz}(R_{\quad xyz}^{0}+R_{\quad zxy}^{0}+R_{\quad
yzx}^{0}).  \label{eq23}
\end{equation}%
There exist the products of two Christoffel symbols, {\it i.e.},
$g^{\sigma \tau }(\Gamma _{\tau ,\alpha \gamma }\Gamma _{\lambda
,\sigma \beta }-\Gamma _{\tau ,\alpha \beta }\Gamma _{\lambda
,\sigma \gamma })$ in the definition of the Riemann curvature,
$R_{\lambda \alpha \beta \gamma }$. Apparently, the products of
two Christoffel symbols ( {\it i.e.}, the nonlinear terms of field
equation ) contain the total indices, $x,y,z$ of three-dimensional
space coordinate ( namely, these indices are taken the
permutations of $x,y,z$ ) and therefore vanish, in the light of
the fact that the Christoffel symbol with index $z$ is vanishing
according to Eq. (\ref{eq21}).

In view of the above discussion, one can conclude that Eq. (\ref{eq22}) can
be exactly reduced to a linear equation. It is easily verified that $%
R_{\lambda \alpha \beta \gamma }$ $(\lambda =x,y,z)$ vanishes with the help
of the linear expression for $R_{\lambda \alpha \beta \gamma }$ given by $%
R_{\lambda \alpha \beta \gamma }=\frac{1}{2}\left({\frac{\partial
^{2}g_{\lambda \gamma }}{\partial x^{\alpha }\partial x^{\beta
}}+\frac{\partial
^{2}g_{\alpha \beta }}{\partial x^{\lambda }\partial x^{\gamma }}-\frac{%
\partial ^{2}g_{\lambda \beta }}{\partial x^{\alpha }\partial x^{\gamma }}-%
\frac{\partial ^{2}g_{\alpha \gamma }}{\partial x^{\lambda
}\partial x^{\beta }}}\right)$ and the linear element expressed by
Eq. (\ref{eq20}). We thus obtain that $R_{\quad \alpha \beta
\gamma }^{0}=g^{00}R_{0\alpha \beta \gamma }.$ By the aid of the
following expression
\begin{equation}
R_{0\alpha \beta \gamma }=\frac{1}{2}\frac{\partial }{\partial
x^{\alpha }}\left({
\frac{\partial g_{0\gamma }}{\partial x^{\beta }}-\frac{\partial g_{0\beta }%
}{\partial x^{\gamma }}}\right),  \label{eq24}
\end{equation}%
one can arrive at
\begin{equation}
\epsilon ^{0\alpha \beta \gamma }R_{\quad \alpha \beta \gamma }^{0}=-\frac{%
g^{00}}{\sqrt{-g}}\nabla \cdot (\nabla \times {\bf g}),  \label{eq25}
\end{equation}%
where the gravitomagnetic vector potentials, ${\bf g,}$ are defined to be $%
{\bf g}=(-g_{0x},-g_{0y},-g_{0z})$. Substitution of Eq. (\ref{eq25}) into
Eq. (\ref{eq22}) yields
\begin{equation}
\nabla \cdot (\nabla \times {\bf g})=-\frac{\sqrt{-g}}{g^{00}}\rho _{M}.
\label{eq26}
\end{equation}%
Note that Eq. (\ref{eq26}) is the exact static gravitational field equation
of gravitomagnetic matter derived from Eq. (\ref{eq13}), where use is made
of the expression (\ref{eq20}) of linear element. For the case of
cylindrically symmetric gravitomagnetic field with the nonvanishing $\rho
_{M}$ being present only in the $x$-$y$ plane with $z=0$, the
gravitomagnetic field, ${\bf B}_{g}$, which is defined as $\nabla \times
{\bf g},$ may be written
\begin{equation}
({\bf B}_{g})_{z}=-\frac{\sigma
_{M}}{2}\left({\frac{\sqrt{-g}}{g^{00}}}\right)_{z=0} \label{eq27}
\end{equation}%
with $\sigma _{M}$ being the surface density associated with $\rho _{M}$ and
$\left({\frac{\sqrt{-g}}{g^{00}}}\right)_{z=0}$ denoting the value of $\frac{\sqrt{-g}}{%
g^{00}}$ in the $x$-$y$ plane where $z=0.$ It follows from Eq. (\ref{eq27})
that the direction of ${\bf B}_{g}$ is parallel to the $z$-axis. The metric
components, $g_{0x},g_{0y},$ are therefore readily obtained as follows
\begin{equation}
g_{0x}=\frac{B_{g}}{2}y,\quad g_{0y}=-\frac{B_{g}}{2}x,  \label{eq28}
\end{equation}%
with $B_{g}=-\frac{\sigma
_{M}}{2}\left({\frac{\sqrt{-g}}{g^{00}}}\right)_{z=0}.$

In order to obtain the contravariant metric $g^{\mu \nu },$ we calculate the
inverse matrix of the metric $\left( g_{\mu \nu }\right) $ and the result is
given as follows
\begin{equation}
\left( g^{\mu \nu }\right) =\frac{1}{1+g_{0x}^{2}+g_{0y}^{2}}\left(
\begin{array}{cccc}
1 & g_{0x} & g_{0y} & 0 \\
g_{0x} & -(1+g_{0y}^{2}) & g_{0x}g_{0y} & 0 \\
g_{0y} & g_{0x}g_{0y} & -(1+g_{0x}^{2}) & 0 \\
0 & 0 & 0 & -(1+g_{0x}^{2}+g_{0y}^{2})%
\end{array}%
\right) .  \label{eq29}
\end{equation}

It is well known that a photon propagating inside the noncoplanarly curved
optical fiber that is wound smoothly on a large enough diameter\cite%
{Chiao,Tomita,Kwiat} is acted upon by an effective Lorentz force which may
be expressed as (in the unit $\hbar =c=1$)\cite{Shen3}
\begin{equation}
{\bf f\equiv }\stackrel{\cdot }{{\bf k}}={\bf k}\times \left( \frac{\stackrel%
{\cdot }{{\bf k}}\times {\bf k}}{k^{2}}\right)  \label{eq30}
\end{equation}%
where the effective magnetic field is $\frac{\stackrel{\cdot }{{\bf k}}%
\times {\bf k}}{k^{2}}$ with dot denoting the time rate of change of ${\bf k}%
(t).$ Eq. (\ref{eq19}) has been shown to be the expression for the
gravitational force acting on the gravitomagnetic monopole, in the similar
manner, one can consider the motion of gravitoelectric charge, for instance,
a photon propagating in the static cylindrically symmetric gravitomagnetic
field. For the sake of analyzing this problem conveniently, we first take
into account the time evolution of wavefunction of a photon in the weak
gravitational field. The infinitesimal rotation operator of wavefunction of
the photon in gravitomagnetic field is given by $U_{R}=1-i\Delta {\bf %
\vartheta }\cdot {\bf J}$ with\cite{Li1}

\begin{equation}
\Delta {\bf \vartheta }=\frac{{\bf k}(t)\times \left[ {\bf k}(t)+\Delta {\bf %
k}\right] }{k^{2}}=\frac{{\bf k}(t)\times \stackrel{\cdot }{{\bf k}}(t)}{%
k^{2}}\Delta t,  \label{eq300}
\end{equation}%
where $\Delta {\bf k}$ is defined to be $\stackrel{\cdot }{{\bf k}}$ $\Delta
t$. Simple calculation shows that the effective Hamiltonian is of the form $%
H_{eff}=\frac{\stackrel{\cdot }{{\bf k}}\times {\bf k}}{k^{2}}\cdot {\bf J}.$
Given that this infinitesimal rotation of wavefunction of the photon is
caused by the gravitomagnetic field, it follows from both Eq. (\ref{eq30})
and (\ref{eq300}) that the equation of motion of a photon in gravitomagnetic
field may be written in the following form
\begin{equation}
\frac{\stackrel{\cdot }{{\bf k}}\times {\bf k}}{k^{2}}=\nabla \times {\bf g}.
\end{equation}%
This formulation is readily generalized to the case of massive particle
moving in gravitomagnetic field and a number of related topics concerning
the matter wave in gravitomagnetic field may be further investigated.

\section{A potential interpretation of anomalous attractive force acting on
Pioneer spacecrafts}

Taking the effects of gravitomagnetic charge into consideration is believed
to be of essential significance in resolving some problems and paradoxes. An
illustrative example that has been briefly discussed is its application to
the problem of cosmological constant. Additionally, in 1998, Anderson {\it %
et al.} reported that, by ruling out a number of nongravitational
potential causes such as the solar radiation pressure,
precessional attitude-control maneuvers, nonisotropic thermal
radiation, radiation of the spacecraft radio beam and so on, radio
metric data from the Pioneer 10/11, Galileo and Ulysses
spacecrafts indicate an apparent anomalous, constant, acceleration
acting on the spacecraft with a magnitude $\sim 8.5\times $ $10^{-8}$cm/s$%
^{2}$ directed towards the Sun\cite{Anderson}. Is it the effects of dark
matter or a modification of gravity? Unfortunately, neither easily works. It
is interesting that, by taking the cosmic mass, $M=10^{53}$ kg, and cosmic
length scale, $R=10^{26}$ m, our calculation shows that this anomalous
acceleration is just equal to the value of gravitational field strength on
the cosmic boundary due to the total cosmic mass. This fact leads us to
consider a theoretical mechanism to interpret this anomalous phenomenon. The
author favors that the gravitational Meissner effect may serve as a
potential interpretation. Here we give a rough analysis, which contains only
the most important features rather than the precise details of this
theoretical explanation. Once gravitomagnetic matter exists in the universe,
parallel to London$^{,}$s electrodynamics of superconductivity,
gravitational field may give rise to an {\sl effective mass }$m_{g}=\frac{%
\hbar }{c^{2}}\sqrt{8\pi G\rho _{m}}$ due to the self-induced charge current%
\cite{Hou}, where $\rho _{m}$ is the mass density of the universe (For the
case of arbitrarily strong gravitational fields, see the interesting work of
Visser\cite{Visser}, where he defined the mass term of graviton by
introducing a non-dynamical background metric). A constant acceleration, $a,$
may result from the Yukawa potential and can be written as
\begin{eqnarray}
a &=&\frac{GM}{2}\left( \frac{m_{g}c}{\hbar }\right) ^{2}=\frac{GM}{c^{2}}%
(4\pi \rho _{m}G)  \nonumber \\
&=&\frac{GM}{c^{2}R}\frac{G(4\pi R^{3}\rho _{m})}{R^{2}}\simeq \frac{GM}{%
R^{2}},  \label{eq33}
\end{eqnarray}%
where for the universe, use is made of $\frac{GM}{c^{2}R}\simeq
{\mathcal O}(1), \quad 4\pi R^{3}\rho _{m}\simeq M$, which holds
when approximate estimation is performed. Note, however, that this
is an acceleration of repulsive force directed, roughly speaking,
from the center of the universe. By analyzing the NASA$^{,}$s
Viking ranging data, Anderson, Laing, Lau {\it et al.} concluded
that the anomalous acceleration does not act on the body of large
mass such as the Earth and Mars. If gravitational Meissner effect
only affected the gravitating body of large mass or large scale
rather than spacecraft, then seen from the Sun or Earth, there
exists an added attractive force acting on the spacecraft. In the
following, ${\bf g}_{\rm free}$ denotes the free-body acceleration
caused by the cosmic mass, and apparently the velocity of the Sun
and the spacecraft agrees with the following equation (using
approximate analysis)
\begin{equation}
\frac{{\rm d}}{{\rm d}t}{\bf v}_{\rm sun}=-{\bf g}_{\rm free}+{\bf a},\quad \frac{%
{\rm d}}{{\rm d}t}{\bf v}_{\rm spacecraft}=-{\bf g}_{\rm free},
\label{eq34}
\end{equation}%
then the velocity of the spacecraft relative to the Sun satisfies
\begin{equation}
\frac{{\rm d}}{{\rm d}t}\left({{\bf v}_{\rm spacecraft}-{\bf
v}_{\rm sun}}\right)=-{\bf a}, \eqnum{35}  \label{eq35}
\end{equation}%
where the acceleration of spacecraft due to the Solar gravitation has been
ignored. The negative sign on the right handed side of Eq. (\ref{eq35})
shows that this added force gives rise to an anomalous, constant,
acceleration directed towards the Sun.

Anderson {\it et al.} reported that without using the apparent
acceleration, {\it i.e.}, using an independent analysis, the
compact high accuracy satellite motion program (CHASMP) shows a
steady frequency drift of about $-6\times 10^{-9}$ Hz/s, or $1.5$
Hz over 8 year. This equates to a clock acceleration, $a_{t}$ of
$-2.8\times 10^{-18}$ s$^{2}$/s. If there were a systematic drift
in the atomic clocks of the DSN (Deep Space Network) or in the
time-reference standard signals, this would appear like a
nonuniformity of time; {\it i.e.}, all clocks would be changing
with a constant acceleration. Anderson {\it et al.} believed that
people cannot rule out this possibility without adequate analysis.
However, we think that the so-called clock acceleration has no
fundamental significance, but favors the existence of the
anomalous acceleration acting on these spacecraft, seen from the
Sun. The relation between coordinate time ${\rm d}x^{0}$ in the
solar system and the proper time ${\rm d}\tau $ is
\begin{equation}
\frac{{\rm d}x^{0}}{c{\rm d}\tau }=1-\frac{\phi _{\rm
sun}+ar}{c^{2}} \label{eq36}
\end{equation}%
with $\phi _{\rm sun}$ being the gravitational potential caused by
the mass of the Sun and $r$ the radial distance from the Sun. By
ignoring $\frac{\phi _{\rm sun}}{c^{2}}$ that is familiar to us,
the added clock acceleration may be written as
\begin{equation}
a_{t}=\frac{{\rm d}^{2}x^{0}}{c{\rm d}\tau ^{2}}=-\frac{a}{c^{2}}\frac{{\rm d%
}r}{{\rm d}\tau }=-\frac{a}{c},  \label{eq37}
\end{equation}%
where $\frac{{\rm d}r}{{\rm d}\tau }$ is taken to be $c,$ the speed of the
radio signals from the spacecrafts. It follows from Eq. (\ref{eq37}) that
\begin{equation}
a=-ca_{t}.  \label{eq38}
\end{equation}%
This expression can be proved correct by inserting the experimental values, $%
a_{t}=-2.8\times 10^{-18}$ s$^{2}$/s and $a=8.5\times $ $10^{-10}$m/s$^{2},$
into the above equation.

Although it seems natural to interpret this anomalous gravitational
phenomenon, there is still something that deserves further research. Why
does not the gravitational Meissner effect explicitly affect the body of
small mass? Maybe small-mass flow cannot serve as the self-induced charge
current that provides the gravitational field with an effective mass in the
field equation. The theoretical resolution of this problem is not very
definite at present. However, the sole reason that the above resolution of
the anomalous acceleration is somewhat satisfactory lies in that no
adjustable parameters exist in this theoretical mechanism. It is one of the
most important advantages in the above mechanism of gravitational Meissner
effect, compared with some possible theories of modification of gravity\cite%
{Nieto}, which are always involving several parameters that cannot
be determined by theory itself. These theories of modification of
gravity were applied to the problem of the anomalous acceleration
but could not calculate the value of the anomalous acceleration
consistent with experiments performed by Anderson {\it et al.} .
In spite of some defects, the theoretical mechanism of
gravitational Meissner effect suggested above still gives a
valuable insight into the problem of anomalous acceleration acting
on the Pioneer spacecrafts.

\section{Non-analytic property of metric and topological gravitomagnetic
charge}

Since the gravitomagnetic charge is topological charge of spacetime, it may
lead to the non-analytic property of space-time metric, and result in the
violation of Ricci identity. This may be illustrated as follows:

Based on the gravitational field equation (\ref{eq11}) of gravitomagnetic
charge, we calculate the following formula%
\begin{equation}
\epsilon _{\alpha \beta }^{\quad \ \mu \nu }\tilde{G}_{\mu \nu }=\epsilon
_{\alpha \beta }^{\quad \ \mu \nu }(\kappa _{1}K_{\mu \nu }+\kappa
_{2}H_{\mu \nu }),  \label{eq39}
\end{equation}%
and obtain%
\begin{eqnarray}
\epsilon _{\alpha \beta }^{\quad \ \mu \nu }(\kappa _{1}K_{\mu \nu }+\kappa
_{2}H_{\mu \nu }) &=&\kappa _{1}H_{\alpha \beta }+4\kappa _{2}K_{\alpha
\beta },  \nonumber \\
\epsilon _{\alpha \beta }^{\quad \ \mu \nu }\tilde{G}_{\mu \nu }
&=&2(R_{\alpha \beta }-R_{\beta \alpha }),  \label{eq40}
\end{eqnarray}%
or%
\begin{equation}
R_{\alpha \beta }-R_{\beta \alpha }=\frac{1}{2}\kappa _{1}H_{\alpha \beta
}+2\kappa _{2}K_{\alpha \beta },  \label{eq41}
\end{equation}%
where use is made of $\epsilon _{\alpha \beta \nu \mu }\epsilon ^{\lambda
\sigma \tau \mu }=g_{[\alpha }^{[\lambda }g_{\beta }^{\sigma }g_{\nu
]}^{\tau ]}$ with $\lambda ,\sigma ,\tau $ and $\alpha ,\beta ,\nu $ being
completely antisymmetric, respectively. It follows from Eq. (\ref{eq11}) and
(\ref{eq41}) that%
\begin{eqnarray}
2\kappa _{2}(R_{\alpha \beta }-R_{\beta \alpha })-\kappa _{1}\tilde{G}%
_{\alpha \beta } &=&(4\kappa _{2}^{2}-\kappa _{1}^{2})K_{\alpha \beta },
\nonumber \\
2\kappa _{1}(R_{\alpha \beta }-R_{\beta \alpha })-4\kappa _{2}\tilde{G}%
_{\alpha \beta } &=&(\kappa _{1}^{2}-4\kappa _{2}^{2})H_{\alpha \beta }.
\label{eq46}
\end{eqnarray}%
It is known that $R_{\alpha \beta }=R_{\beta \alpha }$ holds in the absence
of gravitomagnetic charge in general relativity. However, when there exists
gravitomagnetic charge, the symmetric property of Ricci tensor is no longer
valid. Further calculation shows that
\begin{equation}
R_{\alpha \beta }-R_{\beta \alpha }=\frac{1}{2}g^{\sigma \tau }\left({\frac{%
\partial ^{2}g_{\sigma \tau }}{\partial x^{\beta }\partial x^{\alpha }}-%
\frac{\partial ^{2}g_{\sigma \tau }}{\partial x^{\alpha }\partial x^{\beta }}%
}\right),  \label{eq47}
\end{equation}%
and then it is apparently seen that the non-analytic property of metric
leads to the non-integral condition, and that the above case is in exact
analogy with that of the Aharonov-Bohm effect in electrodynamics. This,
therefore, implies that gravitomagnetic charge possesses the topological or
global properties, and that the geometric and non-analytic properties of
spacetime deserve detailed study.

\section{Concluding remarks}

The present paper considers the concept of gravitomagnetic charge
and its dynamics in curved spacetime, containing the field
equation and the equation of motion and the static exact solution.
Differing from the symmetric property (with respect to the
space-time coordinate) of gravitational field equation of
gravitoelectric matter, the field equation of gravitomagnetic
matter possesses the antisymmetric property. This, therefore,
implies that the number of the non-analytic metric functions is no
more than 6. Although we have no observational evidences for the
existence of gravitomagnetic charge, it is still of theoretical
significance to investigate the gravity theory of the topological
dual mass. It is interesting to speculate on the possibility that
the formation mechanism of gravitomagnetic charge may be found in
the theory of gravitational interaction, just as the case of the
magnetic monopole in various gauge interactions. Once it is
present in universe indeed, gravitomagnetic charge leads to
physically interesting consequences in astrophysics and cosmology,
particularly in the physics of early universe. In the
gravitational fields where gravitomagnetic charge exists, the
influence of gravitational Meissner effect on the cosmological
evolution should not be neglected. In the regime of strong gravity
such as the stages of quantum cosmology and inflationary model,
gravitational Meissner effect may cause important effects. From
the point of view of the framework of classical field theory,
matter may be classified into two categories: {\it gravitomagnetic
matter} and {\it gravitoelectric matter}, according to their
different gravitational features\footnote{The latter is referred
to as the {\it ordinary matter} and the former can therefore be
referred to as the {\it dual matter}. }. Although it is the
classical solution to the field equation, this kind of topological
gravitomagnetic monopoles may arise not as fundamental entities in
gravity theory.

With foreseeable improvements in detecting and measuring technology, it is
possible for us to investigate quantum mechanics in weak-gravitational
fields \cite{Lamm,Alvarez} associated with gravitomagnetic fields, and the
investigation of dual mass is therefore essential in gravity theory. A
potential application of gravitational Meissner effect to the problem of the
anomalous acceleration is of interest. Although there exists the curiosity
as to whether the gravitational Meissner effect is universal for all the
gravitating body, regardless of its mass and scale, it gives the theoretical
value of anomalous acceleration which is somewhat consistent with the
experimental value. This theoretical possibility is of use, given that we
have no plausible explanation so far. It is apparently worthwhile to improve
this mechanism based on the gravitational Meissner effect.

Acknowledgments This project is partially supported by the
National Natural Science Foundation of China under the project No.
$30000034$.


\begin{references}
\bibitem{Kleinert} Kleinert, H., Gen. Rel. Grav. {\bf 32,} 1271 (2000).

\bibitem{Shen} Shen, J. Q., Zhu, H. Y. and Li, J., Acta Phys. Sini. {\bf 50}%
,1884 (2001).

\bibitem{Shen2} Shen, J. Q., Zhu, H. Y., Shi, S. L. and Li, J., Phys. Scr.
{\bf 65}, 465 (2002).

\bibitem{Bohm} Aharonov, Y. and Bohm, D., Phys. Rev. {\bf 115, }485 (1959).

\bibitem{Aharonov} Aharonov, Y. and Carmi, G., Found. Phys{\sl .} {\bf 3},
493 (1959).

\bibitem{Anandan} Anandan, J., J. Phys. Rev. D {\bf 15}, 1448 (1977).

\bibitem{Dresden} Dresden, M. and Yang, C. N., Phys. Rev{\sl .} D {\bf 20},
1846 (1979).

\bibitem{Mashhoon1} Mashhoon, B., Gen. Rel. Grav. {\bf 31}, 681 (1999).

\bibitem{Mashhoon2} Mashhoon, B., Class. Quant. Grav. {\bf 17}, 2399 (2000).

\bibitem{Hogan} Ellis, G. F. R. and Hogan, P. A., Gen. Rel. Grav. {\bf 29,}
235 (1997).

\bibitem{Ingraham} Ingraham, R. L., Gen. Rel. Grav. {\bf 29},117 (1997).

\bibitem{Dirac} Dirac, P. A. M., Proc. Roy. Soc. (London) A {\bf 133}, 60
(1931).

\bibitem{Schwinger} Schwinger, J., Phys. Rev{\sl .} {\bf 144}, 1087 (1966).

\bibitem{Yang1} Yang, C. N., Phys. Rev. D {\bf 1}, 2360 (1970).

\bibitem{Yang2} Yang, C. N., Phys. Rev. Lett. {\bf 33}, 445 (1974).

\bibitem{Hooft} Hooft, G. $^{,}$t, Nucl. Phys. B {\bf 79}, 276 (1974).

\bibitem{Tchra} Tchrakian, D. H., and Zimmerschied, F., Phys. Rev. D {\bf 62}%
, 045002-1 (2000).

\bibitem{Taub} Taub, A. H., Ann. Math.  {\bf 53}, 472 (1951).

\bibitem{Newman} Newman, E. T., Tamburino, L. and Unti, T., J. Math. Phys. {\bf 4}, 915 (1963).

\bibitem{Dowker} Dowker, J. S. and Roche, J. A., Proc. Phys. Soc. London {\bf 92}, 1 (1967).

\bibitem{Nouri} Nouri-Zonoz, M. and Lynden-Bell. D., Class. Quant. Grav. {\bf 14}, 3123 (1997).

\bibitem{Dowker2} Dowker, J. S., Gen. Rel. Grav. {\bf 5}, 603 (1974).

\bibitem{Zee} Zee, A., Phys. Rev. Lett. {\bf 55}, 2379 (1985).

\bibitem{Nouri2} Lynden-Bell. D. and Nouri-Zonoz, M., Rev. Mod. Phys. {\bf 70}, 427 (1998).

\bibitem{Anderson} Anderson, J. D., Laing, P. A., Lau, E. L., {\it et al},
Phys. Rev. Lett. {\bf 81}, 2858 (1998).

\bibitem{Li1} Li, H. Z., Global Properties of Simple Physical Systems
(China: Shanghai Scientific \& Technical Publishers) (1998).

\bibitem{Li2} Li, H. Z., Guo, S. H., Xian, D. C. and Wu, Y. S., Acta Phys.
Sini. {\bf 28}, 549 (1979).

\bibitem{Polyakov} Polyakov, A. M., Phys. Lett. B {\bf 59}, 82 (1974).

\bibitem{Polyakov2} Polyakov, A. M., Nucl. Phys. B {\bf 120}, 249 (1974).

\bibitem{Weinberg} Weinberg, S., Rev. Mod. Phys. {\bf 61}, 1 (1989).

\bibitem{Datta} Datta, D. P., Gen. Rel. Grav. {\bf 27}, 341 (1995).

\bibitem{Alvarenga} Alvarenga, F. G. and Lemos, N. A., Gen. Rel. Grav. {\bf %
30}, 681 (1998).

\bibitem{Cap} Capozziello, S. and Lambiase, G., 1999 Gen. Rel. Grav. {\bf 31}%
, 1005

\bibitem{Chiao} Chiao, R. Y. and Wu, Y. S., Phys. Rev. Lett. {\bf 57}, 933
(1986).

\bibitem{Tomita} Tomita, A. and Chiao, R. Y., Phys. Rev. Lett. {\bf 57}, 937
(1986).

\bibitem{Kwiat} Kwiat, P. G. and Chiao, R. Y., Phys. Rev. Lett. {\bf 66},
588 (1991).

\bibitem{Shen3} Shen, J. Q., Zhu, H. Y. and Shi, S. L., Acta Phys. Sini.
{\bf 51}, 536 (2002).

\bibitem{Hou} Hou, B. Y. and Hou, B. Y., Phys. Ener. Fort. Phys. Nucl. {\bf 3%
}, 255 (1979).

\bibitem{Visser} Visser, M., Gen. Rel. Grav. {\bf 30}, 1717 (1998).

\bibitem{Nieto} Nieto, M. M. and Goldman, T., Phys. Rep. {\bf 205}, 221
(1991).

\bibitem{Lamm} Lammerzahl, C., Gen. Rel. Grav. {\bf 28}, 1043 (1996).

\bibitem{Alvarez} Alvarez, C. and Mann, R., Gen. Rel. Grav. {\bf 29}, 245
(1997).
\end{references}
\end{document}